\documentclass{pasa}%

\title[Finding RRLs with SkyMapper]{Finding RR Lyrae Stars with SkyMapper: an Observational Test}
\author[Akhter et al.]{S. Akhter$^1$, G. S. Da Costa$^1$, S. C. Keller$^1$, B. P. Schmidt$^1$, 
M. S. Bessell$^1$, \and P. Tisserand$^1$\\
\affil{$^1$Research School of Astronomy \& Astrophysics, Australian National University, Canberra, ACT 0200, Australia}}
\jid{PASA}
\doi{10.1017/pas.\the\year.xxx}
\jyear{\the\year}
\usepackage[authoryear]{natbib}
\bibpunct{(}{)}{;}{a}{}{,}
\begin{document}%
\begin{abstract}
One of the major science goals of the SkyMapper Survey of the Southern Hemisphere sky is the determination of the shape and extent of the halo of the Galaxy. In this paper we quantify the likely efficiency and completeness of the survey as regards the detection of RR Lyrae variable stars, which are excellent tracers of the halo stellar population. We have accomplished this via observations of the RR Lyrae-rich globular cluster NGC 3201. We find that for single epoch $uvgri$ observations followed by two further epochs of $g, r$ imaging, as per the intended three-epoch survey strategy, we recover known RR Lyraes with a completeness exceeding 90\%. We also investigate boundaries in the gravity-sensitive single-epoch two-color diagram that yield high completeness and high efficiency (i.e., minimal contamination by non-RR Lyraes) and the general usefulness of this diagram in separating populations.
\end{abstract}
\maketitle%
\section{INTRODUCTION}

The formation and evolution of galaxies like our own Milky Way, and the shape and extent of its dark matter halo, are subjects of significant importance to Astrophysics.  In that context, we can improve our knowledge of galaxy formation processes by studying the sub-structures and stellar streams in the Galactic halo. RR Lyrae stars (RRLs) make excellent halo tracers for such studies as they have well established absolute magnitudes and thus measure accurate distances. Moreover, because of their intrinsic luminosity, they are detectable to the edge of the Galaxy with moderate sized telescopes. RRLs are old and metal poor, and are relatively common and easy to recognise from their characteristic colours and light curves.

Various studies have already been performed using RRLs to aid our understanding of galaxy formation. For example, \citet[and references therein]{akhter12} have studied the break radius, i.e., the Galactocentric radius at which the density profile changes slope, in the spatial density distribution of RRLs in the Milky Way stellar halo. Overdensities and sub-structures also provide important information and RRLs have  been used extensively in this area (e.g., \citealt{vivas01,newberg02,majewski03,ibata03,yanny03,zinn04,belokurov06,vivaszinn06,duffau06,newberg07,belokurov07,keller08a,prior09,watkins09,drake13}). RRLs are also helpful in studying the horizontal-branch (HB) morphology of the halo field as a function of Galactocentric radius \citep[e.g.,][]{majewski93}.

Over the past decade data from the Sloan Digital Sky Survey (SDSS) have been used extensively to identify a number of Galactic halo sub-structures through the use of RRLs \citep[e.g.,][]{watkins09}. \citet{ivezic05} provided selection criteria to identify candidate RRLs from single epoch data using colours generated from the five filter set of the SDSS. Though the SDSS and its extensions have a few strips that reach into the southern hemisphere, the study of RRLs in this part of the sky is currently incomplete. However, the SkyMapper telescope, which will perform a deep digital survey of the southern sky (\citealt{keller07}) using a six filter set
 that includes an extra filter ($v$) compared to the 
 SDSS\footnote{The SkyMapper filter band passes and their relation to those used in the SDSS 
 are discussed in
 \citet{bessell11}.}, promises to redress the situation. The first data release will include three epoch data for two filters, $g$ and $r$, and data taken at a single epoch for the other filters ($u$, $v$, $i$ and $z$). The likely difference between the first two epochs of the $g,r$ imaging will be $\sim$4 hours, while the second and the third epoch observations will be taken with 1-3 days difference (\citealt{keller07}). The multi-epoch data is intended to find variables and here we investigate their usefulness for detecting RRLs.  The SkyMapper filter set is also designed to provide stellar surface gravity information.  Consequently, we also investigate the utility of the gravity-sensitive diagram for separating genuine RRL candidates from potential main sequence contaminants such as eclipsing binaries.  

This paper demonstrates a method to find RRLs in the SkyMapper survey data and investigates the efficiency and completeness by studying an RRL-rich globular cluster. Section~\ref{sm3.2} describes the observations and the data reduction, including the photometric calibration procedure. In Section~\ref{sm3.3}, we outline the RRL selection process which uses variability in the $g$ and $r$ magnitudes in the multi-epoch data together with other colour information. The outcome of our variable selection process in comparison with the known variables in the cluster field is examined in Section~\ref{sm3.4}. In Section~\ref{sm3.5}, we quantify the efficiency and completeness of the RRL selection process. Finally, we discuss our results in Section~\ref{sm3.6}.

\section{OBSERVATIONS AND DATA REDUCTION}\label{sm3.2}

\subsection{The Data}\label{sm3.2.1}

The RRL-rich Milky Way globular cluster NGC~3201 was selected as a test object to measure the utility of the SkyMapper Survey in identifying RRL stars. Key parameters for the cluster are provided in Table ~\ref{table:ngc3201} where ``$r_{core}$" and ``$r_{h}$" are the core radius and the half-light radius in arcmin. The values in the table are taken from the current on-line version of the \citet{harris96} catalogue\footnote{physwww.physics.mcmaster.ca/$\sim$harris/mwgc.dat} except for the number of RRLs in the cluster field which comes from the on-line version of the Globular Cluster Variable Star Catalogue \citep{clement01}\footnote{www.astro.utoronto.ca/$\sim$cclement/read.html}.

\begin{table*}
\caption{Parameters for the Globular Cluster NGC 3201$^a$} 
\label{table:ngc3201}
\begin{center}
\begin{tabular*}{\textwidth}{ccccccccc}
\hline\hline
 ID  & R.A.  & Decl.  & $V_{\rm{HB}}$  & No. of & [Fe/H]  & $E(B-V)$  & $r_{\rm{core}}$  & $r_{\rm{h}}$\\
 & (J2000.0) & (J2000.0) & & RRLs & & & (arcmin) & (arcmin) \\
\hline
NGC 3201 & 10 17 36.82 & $-$46 24 44.9 & 14.76 & 73 & $-$1.59 & 0.24 & 1.30 & 3.10\\
\hline\hline
\end{tabular*}
\end{center}
\medskip
$^a$Entries are from the current online version of the \citet{harris96} catalogue except for the number of RRLs which comes from the online version of the \citet{clement01} catalogue.
\end{table*}

We reproduced the intended three-epoch survey strategy by observing a field centred approximately on the cluster three times.   The first two epochs were separated by 1.93 hours and the second and the third epochs separated by 2.99 days.  Observations were made using the $g$ and $r$ filters at all three epochs and with the $u$, $v$, and $i$ filters at the first epoch only.

The SkyMapper field-of-view is 5.7 deg$^2$ and is imaged by 32 2k$\times$4k CCDs at a scale of 0.5 arcsec per pixel.  Each CCD is readout through 2 amplifiers.  The exposure times were 160 seconds for the $u$ and $v$ filters and 30 seconds for the other filters. All the observations were taken at an airmass $< 1.2$ in the period 21--24 February 2012 during dark time. The image FWHM varied from $\sim$3.3$-$3.8 arcsec for the $g$, $r$, and $i$ filters to $\sim$4.1$-$4.5 arcsec for the $u$ and $v$ filters. The data were processed (bias subtracted, flat-fielded and amp-combined) with a preliminary version of the SkyMapper  Science Data Pipeline System at the ANU Supercomputing Facility, Canberra.

\subsection{Data Analysis}\label{sm3.2.2}

We first separated each frame into the 32 individual CCD images.  These were then fit with a World Coordinate System (wcs) using custom SkyMapper software (Keller 2012, priv.\ comm.).  The cluster was centred on CCD-31 and since the $34^\prime \times 17^\prime$ field of this CCD more than encompasses the cluster, subsequent analysis used only the frames from this CCD\@.  The $u$ filter image was chosen to define the stellar sample, which contains both cluster and field stars, as this image has the brightest limiting magnitude and thus stars detected on this frame are likely to be common to all the other frames. Use of IRAF task $daofind$ on the $u$ frame, with a detection threshold set to maximise the number of stars and minimise the number of spurious detections, resulted in the detection of 817 stars.  The same stars were then identified on other frames by using the IRAF tasks $xy2sky$ and $sky2xy$. 

We then calculated for each star the radial distance $r_{\rm{d}}$ from the center of the cluster on the $g$ first epoch frame, which we denote by $g1$ (more generally we refer to frames by ``$fn$'' where $f$ is the filter and $n$ is the epoch). Visual inspection of the $g$ frames showed that within a radius $r_{\rm{d}} < 300$ pixels, i.e., 150 arcsec, image crowding and blending was significant, while outside of this radius the majority of the stellar images were relatively isolated.  Consequently, to minimise the effects of crowding on the photometry, the stars within a radius of 300 pixels from the cluster centre were excluded from the analysis, leaving a sample of 600 stars.  Aperture photometry, using the IRAF task $phot$ and an aperture radius of 4 pixels, was then performed on each frame to generate sets of instrumental magnitudes.  These output lists were then compared to include only those stars with valid measurements on all the data frames.  This left a final sample of 517 stars that will be referred to as the ``all stars'' sample.  

We then corrected the $g2$ and $g3$ magnitudes to the $g1$ instrumental system, and similarly the $r2$ and $r3$ magnitudes to $r1$ system. To do this we plotted  $g2-g1$ and $g3-g1$ against $g1$ and determined the mean magnitude differences from the $g1$ system using $\sim$150 bright stars with $g1 < 16$ mag
($g_{std} \approx 14.2$).
The offsets to bring $g2$ and $g3$ on to the $g1$ system were $0.112 \pm 0.001$ mag and $0.011 \pm 0.001$ mag respectively, where the error is the standard error of the mean. Offsets were determined in a similar fashion for $r2$ and $r3$ as $-0.043 \pm 0.001$ mag and $0.03 \pm 0.001$ mag respectively, using $\sim$180 bright stars with $r1 < 16$ mag ($r_{std} \approx 14.0$).

\subsection{Photometric Calibration}\label{sm3.2.3}

To bring the photometry onto an (approximate) standard system, we first cross-matched the positions of stars in the ``all stars" list with those for Stetson's photometric standard stars in the vicinity of NGC~3201 given in the file {NGC 3201.pho\footnotemark}\footnotetext{www3.cadc-ccda.hia-iha.nrc-cnrc.gc.ca/community/STETSON/\\standards/NGC3201.pho}. This yielded 21 stars in common with $V < 16.0$ and within a 1 arcsec sky position difference limit.  We then carried out a least squares fit to ($V - g1$) versus ($B-V$), where $B$ and $V$ (and also $U$, $R$, and $I$ as used later) are the standard magnitudes from Stetson's photometry, which is on
the \citet{AL92} system.  This resulted in a slope, $\alpha$, and an intercept, $ZP$. The zero point for the SkyMapper photometric system adopted here, denoted by $g_{\rm{std}}$, was then chosen so that:

\begin{equation}
V = g_{\rm{std}}
\end{equation}
for a star with $(B-V)_{0} = 0.0$. Given the reddening of NGC 3201 is $E(B-V) = 0.24$ mag (\citealt{harris96}), we then have the relation:

\begin{equation}
g_{\rm{std}} = g1 + \alpha \times 0.24 + ZP.
\end{equation}
The values of $\alpha$ and $ZP$, along with the $rms$ about the fit, are given in Table~\ref{table:alphazp}.

A similar process was then used to determine offsets to place the instrumental magnitudes $v1$ and $i1$ on a standard system.  As for $g_{\rm{std}}$, the standard magnitudes $v_{\rm{std}}$ and $i_{\rm{std}}$ were calculated from the slopes and intercepts derived from the least squares fits to the magnitude difference-colour relations ($B-v1$) versus ($B-V$) and ($I - i1$) versus ($V-I$), applied at ($B-V$)$_{0}$ = 0.0 and  ($V-I$)$_{0}$ = 0.0, respectively.  We take E($V-I$)/E($B-V$) as 1.25 for ($B-V$)$_{0}$ = 0.0 \citep{dean78} and thus ($V-I$)$_{0}$ = 0.0 corresponds to ($V-I$) = 0.30.  We note further that since the ($B-v1$) versus ($B-V$) relation is somewhat non-linear, the $\alpha$ and $ZP$ values given in Table \ref{table:alphazp} have been determined from the fit to only the bluer stars ($B-V < 0.8$).

Unfortunately, there are no Stetson standard stars in the NGC~3201 field with $U$ and $R$ magnitudes. Consequently, we used the magnitude difference-colour relations ($B-u1$) versus ($B-V$) and ($V-r1$) versus ($V-I$) to determine the offsets to generate standardised Skymapper magnitudes $u_{\rm{std}}$ and $r_{\rm{std}}$. The process works well for the $r$ magnitudes, as demonstrated by the low $rms$ for the fit given in Table \ref{table:alphazp}.  However, even restricting the fit to the bluer stars, the ($B-u1$) versus ($B-V$) relation shows significant scatter, resulting in a larger $rms$ value compared to the other  relations and thus larger uncertainties in the $\alpha$ and $ZP$ values.  Consequently, while we assert that the calibrations for $g$,  $i$, $r$ and $v$ are relatively reliable, the standardised $u$ magnitudes derived here may not be exactly on the final SkyMapper photometric system.  Consequently, the magnitudes and colours discussed in this work should be considered as illustrative only, although none of the subsequent analysis is affected by our choice of zero points.

\begin{table}
\caption{The calculated slopes and zero points adopted for photometric calibration} \label{table:alphazp}
\begin{center}
\begin{tabular}{ccrc}
\hline\hline
Mag diff vs Colour & RMS & Slope & Intercept \\
& (mag) & ($\alpha$) & ($ZP$)\\
\hline
($V-g1$) vs ($B-V$) & 0.019 & $-0.309$ & $-1.730$\\
($B-v1$) vs ($B-V$) & 0.048 & $-0.732$ & $-4.404$\\
($B-u1$) vs ($B-V$) & 0.193 & $-0.551$ & $-5.417$\\
($V-r1$) vs ($V-I$) & 0.015 & 0.330 & $-2.128$\\
($I-i1$) vs ($V-I$)  & 0.015 & $-0.095$ & $-3.085$\\
\hline\hline
\end{tabular}
\medskip\\
\end{center}
\end{table}


We now proceed to correct our standardised magnitudes for the effects of reddening.  We assume that the cluster value of E($B-V$) = 0.24 applies to all stars in the field whether they are cluster members or not.  Bessell (2012, priv.\ comm.) has used the response functions of the SkyMapper filter system, input spectral energy distributions from model atmospheres, and a standard absorption curve to determine absorption and reddening coefficients for the SkyMapper filter system.  In particular, we have A$_{g}$ = 1.19 A$_{V}$, so that with A$_{V}$ = 3.2 E($B-V$), the absorption correction to the standardised $g$ magnitudes is 0.91 mag.

Similarly, for a  6500 K, log $g = 2.5$ spectrum representative of an RRL star, we have E($g-i$) = 1.45 E($B-V$), and E($u-v$) = 0.24 E($B-V$), with the latter relation applying for E($B-V$) $<$ 0.3, as is the case here.  The reddening corrections to the standardised $g-i$ and $u-v$ colours are then 0.35 and 0.06 mag, respectively.  In the subsequent discussion all standardised magnitudes and colours have been reddening corrected.  We note in passing that for the SkyMapper survey itself, reddening values will be taken from 
standard sources such as \citet[][see also \citet{MZN11} and the references therein]{SFD98}, and that 
most field RRLs will be at sufficiently large distances and heights above the Galactic plane that the 
uncertainties in the reddening corrections will be small.










\section{RR Lyrae Selection Process}\label{sm3.3}

We now describe the process through which the candidate RRLs are selected.  In brief, the first step is identifying stars that appear to vary significantly across the 3 epochs of $g$ and $r$ data.  The second step uses the fact that the $g$ and $r$ data were taken almost simultaneously: as a consequence the variations in the $g$ and $r$ magnitudes for genuine variable stars should be tightly correlated.  The third and final step uses the colour of the candidates and their location in a gravity-sensitive diagram to separate the RRL candidates from other classes of potential variables.

\subsection{Variability}\label{sm3.3.1}

For each of the 517 stars in the ``all stars'' sample we calculated the mean magnitudes from the $g$ and $r$ observations at the 3 epochs, and also the standard deviations of the magnitudes.  The mean magnitude values were then grouped into bins (all stars brighter than $\langle g \rangle$, $\langle r \rangle = 14.0$ mag, then bins of width 0.5 mag) and the mean and standard deviation of the individual 3-epoch standard deviations for the stars in each bin calculated.  We denote these quantities by $\langle \sigma_i \rangle_n$ and $\sigma_n(\sigma_i)$, where $n$ represents the bin number and $i$ the set of stars in each bin.  Then, for each magnitude bin, we removed the stars with 3-epoch standard deviations greater than 3$\times\sigma_n(\sigma_i)$, flagging them as candidate variables.  The values of $\langle \sigma_n \rangle$ and $\sigma_n(\sigma_i)$ are then recalculated and further ``3$\sigma$'' deviants classed as candidate variables.  The process is continued until $\sigma_n(\sigma_i)$ ceases to change significantly.

\begin{figure}
\begin{center}
\includegraphics[width=0.42\textwidth]{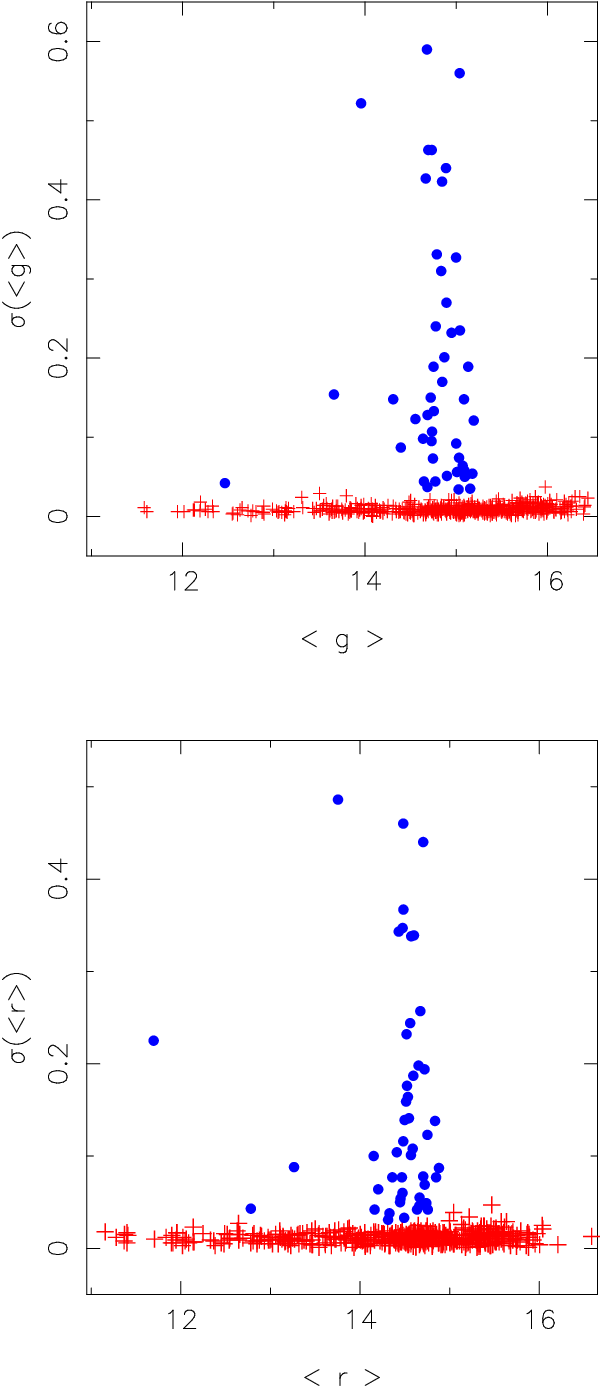}
\caption{The standard deviation of the magnitude values from the 3 observation epochs is shown as a function of the mean magnitudes for the $g$ filter (upper panel) and for the $r$ filter (lower panel).  Using an iterative sigma clipping process 47 $g$ and 49 $r$ candidate variables are identified.  These are shown as blue solid circles while the non-variables are plotted as red plus-signs. }
\label{fig:sm3.1}
\end{center}
\end{figure}

By this process we identified 47 stars as candidate variables from the $g$ frames and 49 from the $r$ frames.  In total there are 51 candidates at this stage, of which 45 are in common between the two sets.  The output is illustrated in the panels of Fig.\ \ref{fig:sm3.1} in which the individual 3-epoch standard deviations are plotted against the mean magnitudes for the $g$ (upper panel) and $r$ data (lower panel).  The candidate variables are shown as filled circles while the stars classified as non-variable are represented by plus-signs.

\subsection{$\Delta(g)$ versus $\Delta(r)$} \label{sm3.3.2}
\begin{figure}
\begin{center}
\includegraphics[width=0.42\textwidth]{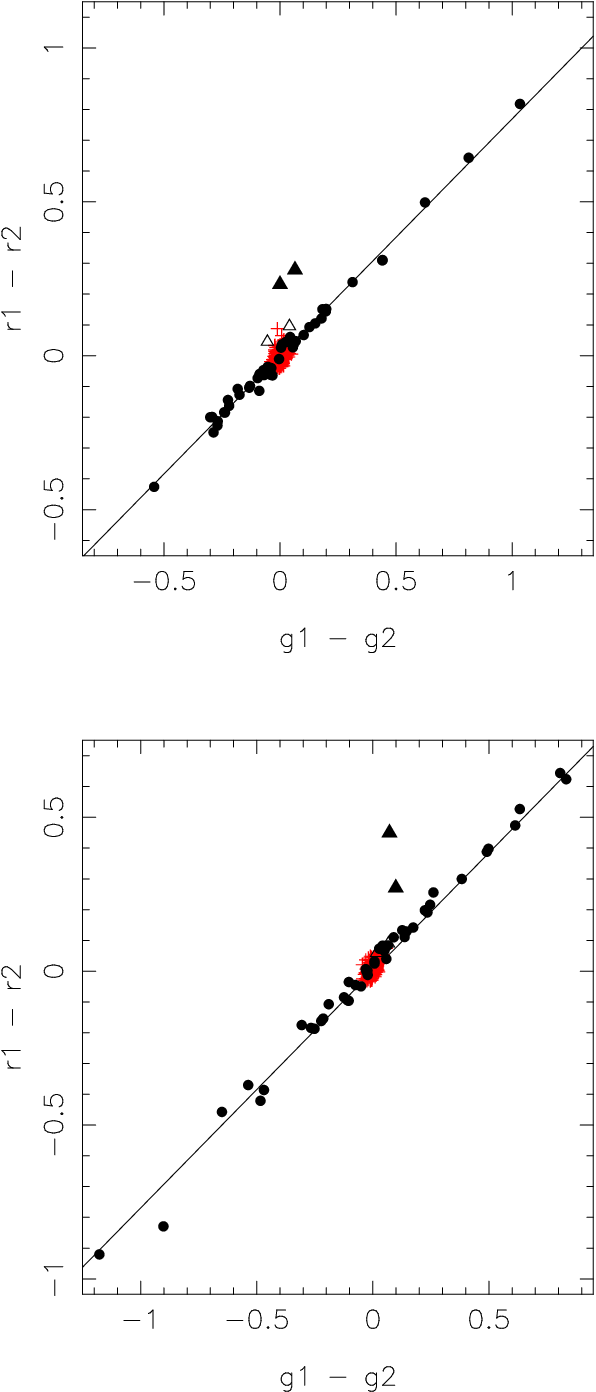}
\caption{Magnitude difference relations for $g$ and $r$ for epochs 1 and 2 (upper panel) and epochs 1 and 3 (lower panel).  The red plus-symbols are stars classified as non-variable from their magnitude dispersion across all 3 epochs (see Fig.\ \ref{fig:sm3.1}), while the black symbols represent the candidate variables.  Most (the filled circles) follow a well defined correlation $\Delta(g)$ = 1.30 $\Delta(r)$, which is shown as the black solid line.  Four stars do not follow this relation and are plotted as triangles -- 2 of the four stars (filled triangles) lie off the relation in both panels while the other two (open triangles) are discrepant only in the upper panel.  These four stars have been removed from the candidate variables list.}
\label{fig:sm3.2}
\end{center}
\end{figure}

For stars that are truly variable, the changes in magnitude for a set of observations through different filters should be tightly correlated provided the observations are nearly simultaneous, as is the case for the $g$ and $r$ observations here.  Consequently, we calculated the magnitude differences $r1-r2$ and $g1-g2$ for the first and second epochs, and similarly the first and third epoch magnitude differences $r1-r3$ and $g1-g3$.  The relationships between these magnitude differences are shown in the panels of Fig.\ \ref {fig:sm3.2}.  It is clear that most of the candidate variables identified in the previous section are true variables as they exhibit tight linear relations between the magnitude differences.  We note though that for RRLs, the 
substantial changes in surface gravity that occur at particular phases in the pulsation cycle might result in 
less well-defined relations 
if SkyMapper $u$ or perhaps $v$ magnitude differences are compared to the $r$ magnitude differences for a
similar ensemble of photometric measurements.  It is also possible that the scatter about the $\Delta(g),
\Delta(r)$ relation might be increased if there is a significant metallicity spread among the RRLs in the sample.

In Fig.\ \ref{fig:sm3.2} there are, however, a small number of stars identified as candidate variables by the previous selection process that lie away from  the correlations defined by the majority of the candidate variables.  These stars were inspected individually on the frames and it was concluded that the apparent variations were due to varying amounts of contamination of the aperture photometry by close neighbours. These stars are thus probably not true variables and have been discarded from the candidate variables list.  Indeed their positions do not match 
with any of the catalogued variables in the NGC~3201 field.

As a result of this process and that of the previous section a final list of 47 candidate variables was generated.  
For these stars the best fit line for both ($\Delta(g), \Delta(r)$) data sets indicates a relation $\Delta(g)$ = 1.30 $\Delta(r)$.  A similar relation, $\Delta(B)$ =1.32 $\Delta(R)$, was found for a sample of 119 MACHO RRLs in the LMC \citep{keller08a}.

\subsection{Surface Gravity Selection}\label{sm3.3.3}
Relative to the SDSS, the SkyMapper system possesses an extra filter, $v$, and a modified $u$ filter.  As shown in \citet{keller07} using model atmosphere based calculations, an appropriate colour combination involving these filters, with ($g-i$) as a temperature indicator, can provide surface gravity information \citep[see Fig.\ 11 of][]{keller07}.  We now investigate whether we can use this colour index to distinguish  candidate RRLs (log $g$ $\sim$ 2.5) from potential main-sequence (log $g$ $\sim$ 4) contaminants such as eclipsing binaries.  We first note that of our 47 candidate variables, 3 have notably red colours (first epoch standardised and dereddened ($g-i$) colours exceeding 1.0) and are therefore unlikely to be genuine RRLs.  The other 44 variable candidates have relatively ``blue'' colours, potentially consistent with an RRL classification.  

\begin{figure}
\begin{center}
\includegraphics[width=0.42\textwidth,angle=-90]{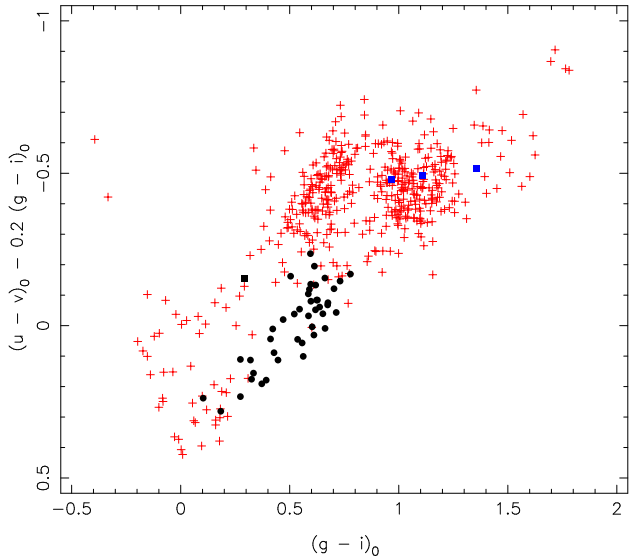}
\caption{The gravity-sensitive index $(u-v)_{0}-0.2(g-i)_{0}$ is plotted against $(g-i)_{0}$ for the ``all stars'' sample.  It is evident that for $(g-i)_{0}$ less than $\sim$0.9 most of the stars fall in one of two distinct branches, which we interpret as indicating the gravity difference between main sequence stars (upper branch) and evolved stars with lower gravities.  Based on this diagram we classify 43 of our candidate variables as RRLs (filled black circles) and one as a main-sequence variable (e.g., an eclipsing binary) shown as the filled black square.  The three remaining variable candidates (blue filled squares) have redder colours and are not readily classified.}
\label{fig:sm3.3}
\end{center}
\end{figure}

In Fig.\ \ref{fig:sm3.3} we plot the gravity-sensitive index $(u-v)_{0}-0.2(g-i)_{0}$ against $(g-i)_{0}$
for the ``all stars'' sample using the epoch 1 photometry.  For the stars bluer than ($g-i$)$_{0}$  $\approx$ 0.9 there is a clear separation into two distinct branches.  Based on Fig.\ 11 of \citet{keller07}\footnote{We note that comparing Fig.\ 11 of \citet{keller07} with our Fig.\  \ref{fig:sm3.3} shows that both the x- and y-axis values are consistent, suggesting that our standardisation and de-reddening procedures are sufficient to validate our interpretation of Fig.\ \ref{fig:sm3.3}.} we interpret the upper branch as being dominated by stars with main sequence gravities while the lower branch is presumed to be dominated by lower gravity evolved stars.  Of the 44 RRL candidates, 43 fall on this lower branch while one variable candidate falls among the upper branch stars.  

We can then summarise the outcome of variable star selection processes as the following: using an observational strategy similar to that for the SkyMapper 3-epoch survey, we have identified 47 candidate variables.  Of these we classify 43 as candidate RRLs, 1 as a candidate main sequence variable, while the remaining 3 are redder objects, which cannot be readily classified.  We now compare this outcome with the catalogue of known variables for this cluster.

\section{COMPARISON WITH KNOWN VARIABLES}\label{sm3.4}

We now proceed to verify our selection processes by comparing our sample of candidate variables with a list of known variables.  For NGC~3201 the most recent catalogue of confirmed variables is given in the on-line file www.astro.utoronto.ca/$\sim$cclement/cat/C1015m461, which draws on data from, for example, \citet{Samus96} and \citet{layden03} (see the Supplementary Notes associated with the file for a full list of references).  The file contains positions for the variables which, with aid of the on-line reference image available at www3.cadc-ccda.hia-iha.nrc-cnrc.gc.ca/
community/STETSON/standards/NGC3201.fits.gz, enabled identification of the variables on the SkyMapper frames.  In our chosen region beyond 150 arcsec from the cluster centre there are 44 RRLs and nine other variables in the catalogue, a list that can be considered as complete, at least as regards RRL stars. 

A cross-match of this list with our candidate variables revealed a match for 42 of our RRL candidates with confirmed RRLs, and a match of our main-sequence candidate variable with a known eclipsing binary (V65) that is not a member of the cluster.   None of our 3 ``red'' candidate variables matched entries in the catalogue.  Among the remaining eight catalogue non-RRL variables, four are fainter than our detection limit while the other four are long period variables ($P$ $\sim$ 11--72 days) which we would not necessarily expect to detect given the limited time span of the three-epoch data. The SkyMapper six epoch data set, which will have additional epochs with approximately one-week, one-month and one-year sampling, will be better suited to detecting variables of this type.

There are two catalogue RRLs that were not detected as candidates, namely V25 and V72.  V25, which 
has period of 0.5148d and $V_{amp}$ = 1.05, is present in the ``all stars'' sample but at the specific epochs of our observations the star did not vary sufficiently to be detected as a variable.  
The situation for V72 is less clear-cut.  In the \citet{clement01} on-line catalogue the star is listed has having a period of 0.36d, but no photometry is given.  We find that there is a star
in our ``all stars'' sample whose position corresponds to that listed for V72 in the on-line catalogue.  
However, our photometry for this star has a 
red colour, $(g-i)_{0}$ = 1.19, and with $g_{0}$ = 12.89 the star is notably brighter than the 
majority of the other cluster RRL candidates.  Further, it shows no evidence of variablity over the 
epochs of our 
observations.   Intriguingly there is a second star in our ``all stars'' sample that has the same right 
ascension as ``V72'' but which lies 12$^{\prime\prime}$ to the south.  This star is fainter ($g_{0}$ = 14.68) 
and possesses a colour much more consistent with those of the RRL candidates ($(g-i)_{0}$ = 0.37).
We suggest therefore that the position for V72 derived in \citet{Samus96} from the original rectangular
coordinates of \cite{Dowes40}, and which is reproduced in the on-line version of the \citet{clement01} 
catalogue, is incorrect and that the actual V72 is the fainter star just south of the
star marked as ``V72'' on the finding chart in \citet{Samus96}.   
The correct position for V72 derived from our analysis
is 10:17:22.24, --46:15:02 (J2000).  Over the epochs of our data the star does not appear to vary.  We note
also that the star lies among the ``upper branch'' stars in the gravity-sensitive diagram shown in 
Fig.\ \ref{fig:sm3.3} suggesting that if it is a variable, it is more likely an eclipsing binary than a RRL
star.  

Consequently, it appears that there are a total of 43 rather than 44 genuine RRLs in the region we have studied, of which we have recovered 42.
On the other hand we have one RRL candidate that is not confirmed as an RRL.  Inspection of the data frames suggests that the photometry of this star has been affected by the presence of a nearby star, resulting in a misclassification as a candidate variable.  The three ``red'' candidate variables are also all false positives with their photometry either biased by close neighbours or by the effects of cosmic rays.


\section{COMPLETENESS OF THE SELECTION PROCESS}\label{sm3.5}

With our three-step selection process we have been able to recover 42 of the 43 known RRL variables in the area of sky studied, with the number of known RRLs considered complete, given the number of previous 
investigations of this cluster \citep[see][and the references therein]{clement01}.  Our results therefore indicate
a {\it RRL detection completeness for the SkyMapper 3-epoch survey exceeding 90 percent}, a very satisfactory outcome.  The false-positive rate (1 from 43) is also reassuringly low at 2--3\%, enhanced by the apparent ability of the gravity-sensitive diagram (Fig.\ \ref{fig:sm3.3}) to discriminate main-sequence variables from genuine RRLs, a task that otherwise requires considerably more epochs \citep[see, e.g.,][]{akhter12}. 



\begin{figure}
\begin{center}
\includegraphics[width=0.42\textwidth]{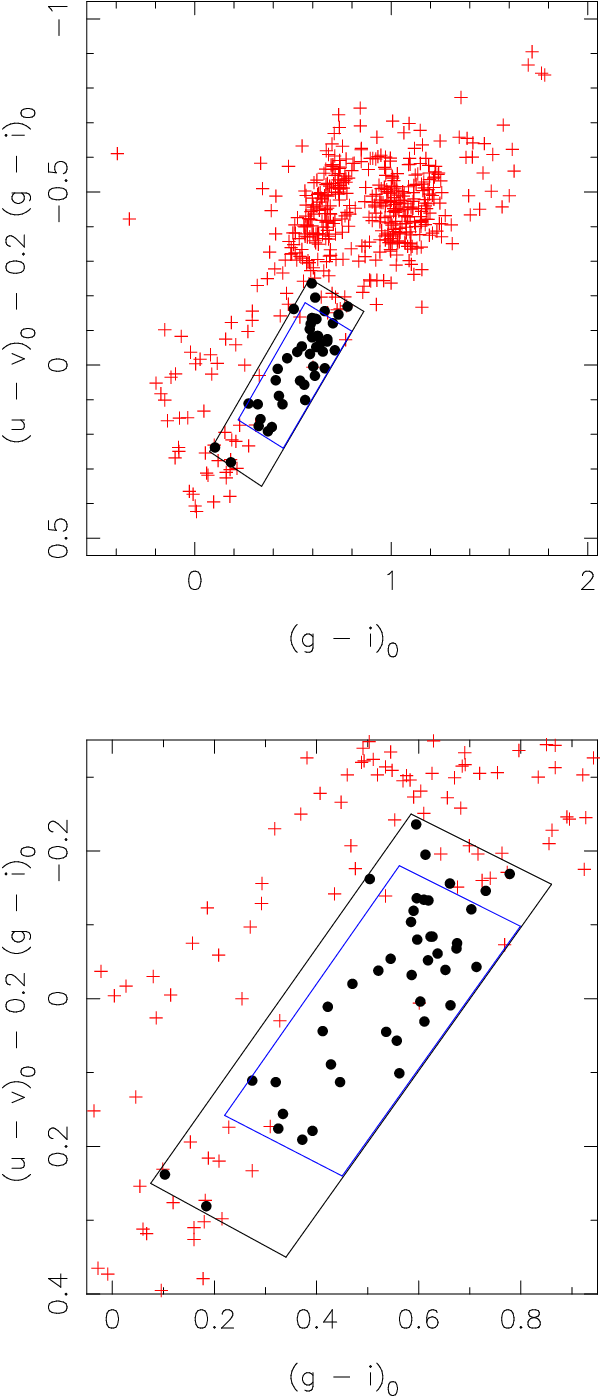}
\caption{The surface gravity plot from Fig.\ \ref{fig:sm3.3} showing the adopted high completeness region, the outer black box, and the high efficiency region, the inner blue box. The lower panel shows an enlarged version of the upper panel.  All of the 43 known RRLs in the region studied are shown as black dots while the other stars from the ``all stars'' sample are shown as red plus-signs.}
\label{fig:sm3.5}
\end{center}
\end{figure}

Now that we have established the outcome of the RRL variable search, we can use the results to quantify how efficiently the gravity-sensitive plot separates RRLs from non-RRLs using solely single-epoch data.  Our approach is similar to that of \citet{ivezic05}.  Specifically, we have selected two regions in the surface-gravity plot, one that has high completeness (C), i.e., encompasses most or all of the RRLs regardless of the contaminant level, and one that has high efficiency (E), i.e., endeavours to maximise the ratio of RRLs to contaminants.  The regions adopted here are illustrated in Figure~\ref{fig:sm3.5}. The region 
outlined by the black box contains all 43 of the known RRLs so that by construction the completeness
is 100\%. There are, however, an additional 19 stars in the region so the efficiency is 69\%.  On the other hand, the region outlined by the blue box contains 35 of the 43 RRLs but only 4 additional stars.  Hence, for this region we can achieve E $\approx$ 90\% for C $\approx$ 80\%.  While illustrative and admittedly for a biased sample (i.e., an RRL-rich globular cluster) these numbers are a distinct improvement compared to those found by \citet{ivezic05} in selecting RRLs from single-epoch SDSS data using the QUEST RRL survey \citep{vivas04}. In that case the selection efficiency found was only 6\% for a 100\% complete colour-selected sample, although it was possible to increase the efficiency up to 60\% for 28\% completeness.  

\section{DISCUSSION}\label{sm3.6}

The RRL selection process studied in this paper used observations with an epoch spacing comparable to those for the SkyMapper 3-epoch survey.  We showed that the combination of variability estimates, correlation of magnitude differences in $g$ and $r$, colour selection, and the use of gravity sensitivity provided by SkyMapper's $u$ and $v$ filters, results in a highly efficient identification of RRL stars -- a completeness above 90\% was achieved.  

However, it is likely that at magnitudes near the faint limit of the survey the completeness may be reduced as the photometry errors will be larger.  Additional follow-up observations may then be required to define the completeness at the faintest magnitudes, which represent the most distant RRLs.  Such observations can nevertheless be executed using modest sized telescopes \citep[e.g.,][]{akhter12}.  
Further, as in
the case of the failure of this particular set of observations to recover V25, the completeness will also depend
to some extent on the exact cadence of the 3-epoch observations, which will vary from field to field, and on the 
period distribution of the RRL stars (e.g., Oo I vs Oo II).  The likely recovery fraction as a function of period is,
however, straightforwardly modeled by Monte-Carlo techniques for a given set of observation epochs.

Overall the results presented here bode well for the full 3-epoch SkyMapper survey.  It will be a rich source of Galactic halo RRLs and thus for studying the shape and extent of the halo and the degree of sub-structure present.   Together with the characteristics of the sub-structures found from the survey data, these properties will provide much new information to improve our understanding of the formation of the Galaxy.

Particularly noteworthy is the apparent ability to distinguish genuine RRLs from main-sequence star contaminants through the use of the gravity sensitive diagram.  Previous surveys, such as the SEKBO RRL survey \citep{keller08a}, which had similar epoch spacing to the SkyMapper 3-epoch survey, have significant contamination from eclipsing binaries -- stars whose characteristics can mimic those of genuine RRLs when only a few observation epochs are available.  \citet{prior09} and \citet{akhter12} estimate that the contamination rate for the SEKBO survey by non-RRL variables is $\sim$25\%; for SkyMapper 3-epoch data it may be
much lower.  The ability to substantially reduce contamination by non-RRL variables without resorting to additional epochs of observation is obviously a big advantage.  Nevertheless, it is important to keep in 
mind that by targetting a globular cluster field for study we are, by deliberate choice, biasing the 
variables detected towards RRLs, whereas in the general field, the relative fraction of other types of 
short-period variables compared to RRLs is larger.  The results shown in Fig.\ \ref{fig:sm3.3} are 
therefore suggestive
but will ultimately require reinforcement from SkyMapper photometry of a larger set of short-period eclipsing
binaries of various types to confirm the utility of the gravity-based RRL selection criteria.

We end by illustrating the general usefulness of the SkyMapper gravity-sensitive diagram in the panels of Fig.\ \ref{fig:sm3.3}.  The upper panel shows the colour-magnitude diagram (CMD) for the entire ``all stars'' sample.  The 43 known RRLs in the region studied are shown as filled symbols and the eclipsing binary
(V65) is shown as the filled square, as is V72.  We note in passing that the RRL at $g_0$ $\sim$ 12.5 in the figure must have been close to maximum light at the epoch of the $g1$ observations, as this star (V81, P=0.5198d, $V_{amp}$=1.13) has a mean magnitude in the NGC~3201 variable star catalogue similar to the other cluster RRLs.  
 
\begin{figure}
\begin{center}
\includegraphics[width=0.42\textwidth]{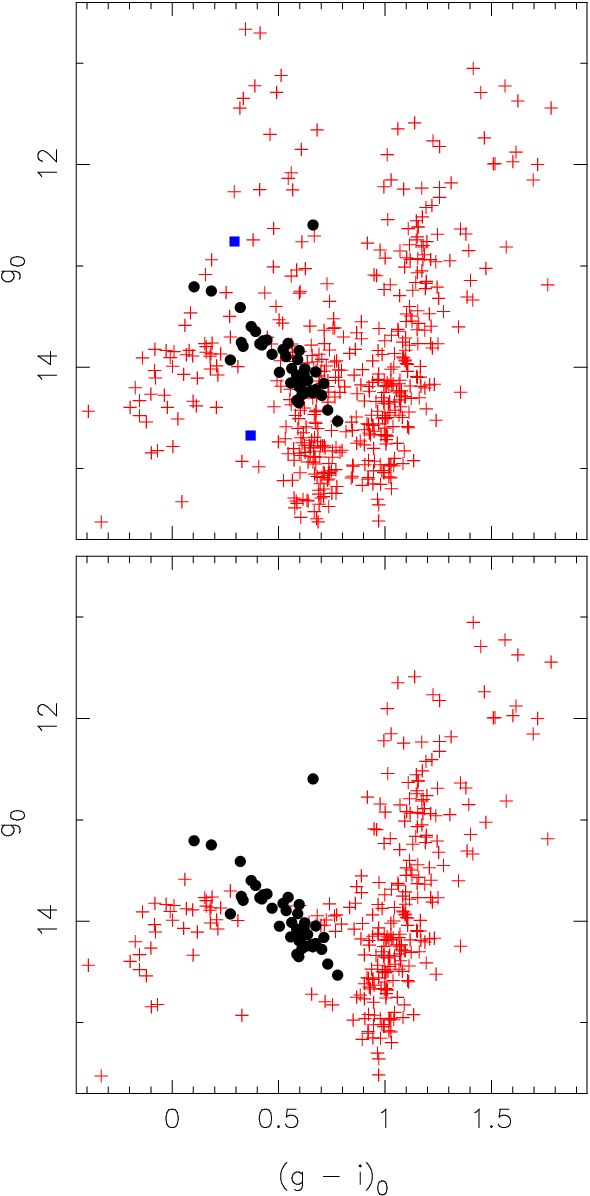}
\caption{The upper panel shows the colour-magnitude diagram for the entire ``all stars" sample.  The 43 known RRLs in the region studied are shown as filled circles while the eclipsing binary (V65) and V72 are show as blue filled squares.  Other stars are plotted as red plus-signs.  The lower panel shows the same data except that the stars with main sequence gravities in Fig.\ \ref{fig:sm3.3} have been excluded.  The horizontal branch of the cluster is now much better defined.}
\label{fig:sm3.4}
\end{center}
\end{figure}

Inspection of Fig.\ \ref{fig:sm3.5} shows that the two ``branches'' are readily separable for ($g-i$)$_0$ less than $\sim$0.9, while Fig.\ 11 of \citet{keller07} shows that the gravity sensitivity is lost at ($g-i$)$_0$ $\sim$ --0.1, as the constant log {\it g} curves turn around and overlap for hotter temperatures.  Consequently, we have plotted in the lower panel of Fig.\ \ref{fig:sm3.4} a CMD where we have eliminated the stars on the upper (main sequence) branch with ($g-i$)$_0$ colours between --0.05 and 0.88 mag.  It is evident that in this process we have removed a large amount of field-star contamination.  The relatively evenly populated horizontal branch (HB) of the cluster\footnote{\citet{lee94} give $(B-R)/(B+V+R)$ = 0.08, for NGC~3201 where B, V and R are the number of blue, variable and red HB stars, respectively; see also the CMD for the cluster in \citet{layden03}.} is now much more evident than it is in the upper panel.  This is also an encouraging outcome as it indicates that the SkyMapper gravity-sensitive diagram can be used to discriminate blue HB stars from higher gravity blue-straggler stars, and to separate red clump stars from higher gravity main sequence stars (at least for ($g-i$)$_0$ $<$ $\sim$0.9).  We note that existing studies of distant blue HB star candidates, for example, generally require spectroscopic observations to separate genuine blue HB stars from nearer blue-straggler contaminants \citep[e.g.,][]{deason12}.  Clearly the SkyMapper survey data will allow the generation of significantly cleaner input catalogues for such studies.

To summarise, we have demonstrated a selection process to search for RRLs on SkyMapper 3-epoch data that has a likely completeness of greater than 90\%. We have also demonstrated the ability of the SkyMapper gravity-sensitive diagram to separate evolved stars from main-sequence objects.  Both will be very valuable techniques to apply when the first SkyMapper data release becomes available.

\begin{acknowledgements} 
We thank the anonymous referee for thoughtful comments that led to improvements in the manuscript.  
SkyMapper operations are funded by the Research School of Astronomy \& Astrophysics of the Australian National University.  Research conducted with SkyMapper is supported in part by grants from the Australian Research Council, most recently through ARC Laureate Fellowship FL0992131 and Discovery Projects grant DP120101237.
\end{acknowledgements}


\begin{thebibliography}{}

\bibitem[Akhter et al.(2012)]{akhter12} Akhter, S., Da Costa, G. S., Keller, S. C., \& Schmidt, B. P. 2012, ApJ, 756, 23
\bibitem[Belokurov et al.(2006)]{belokurov06} Belokurov, V., Zucker, D. B., Evans, N.W., 
et al. 2006, ApJ, 642, L137
\bibitem[Belokurov et al.(2007)]{belokurov07} Belokurov, V., Evans, N. W., Bell, E. F., 
et al. 2007, ApJ, 657, L89
\bibitem[Bessell et al.(2011)]{bessell11} Bessell, M., Bloxham, G., Schmidt, B., et al.\ 2011, PASP, 123, 789 
\bibitem[Clement et al.(2001)]{clement01} Clement, C., Muzzin, A., Dufton, Q., et al.\ 2001, AJ, 122, 2587
\bibitem[Dean et al.(1978)]{dean78} Dean, J. F., Warren, P. R.,  \& Cousins, A. W. J. 1978, MNRAS, 183, 569
\bibitem[Deason et al.(2012)]{deason12} Deason, A. J., Belokurov, V., Evans, N. W., et al. 2012, MNRAS, 425, 2840
\bibitem[Drake et al.(2013)]{drake13} Drake, A. J., Catelan, M., Djorgovski, S. G., et al. 2013, ApJ, 763, 32
\bibitem[Dowse(1940)]{Dowes40} Dowse, M. 1940, Harvard Obs.\ Bull., no.\ 913, p.\ 17
\bibitem[Duffau et al.(2006)]{duffau06} Duffau, S., Zinn, R., Vivas, A. K., Carraro, G., M{\'e}ndez, R. A., Winnick, R., \& Gallart, C. 2006, ApJ, 636, L97
\bibitem[Harris(1996)]{harris96} Harris, W. E. 1996, AJ, 112, 1487
\bibitem[Ibata et al.(2003)]{ibata03} Ibata, R. A., Irwin, M. J., Lewis, G. F., Ferguson, A. M. N., \& Tanvir, N. 2003, MNRAS, 340, L21
\bibitem[Ivezi{\'c} et al.(2005)]{ivezic05} Ivezi{\'c}, {\v Z}., Vivas, A. K., Lupton, R. H., \& Zinn, R. 2005, AJ, 129, 1096
\bibitem[Keller et al.(2007)]{keller07} Keller, S. C., Schmidt, B. P., Bessell, M. S., et al. 2007, PASA, 24, 1
\bibitem[Keller et al.(2008)]{keller08a} Keller, S. C., Murphy, S., Prior, S., Da Costa, G., \& Schmidt, B. 2008, ApJ, 678, 851
\bibitem[Landolt(1992)]{AL92} Landolt, A. U. 1992, AJ, 104, 340
\bibitem[Layden \& Sarajedini(2003)]{layden03} Layden, A. C., \& Sarajedini, A. 2003, AJ, 125, 208
\bibitem[Lee et al.(1994)]{lee94} Lee, Y.-W., Demarque, P., \& Zinn, R. 1994, ApJ, 423, 248
\bibitem[Majewski(1993)]{majewski93} Majewski, S. R. 1993, ARA\&A, 31, 575
\bibitem[Majewski et al.(2003)]{majewski03} Majewski, S. R., Skrutskie, M. F., Weinberg, M. D., \& Ostheimer, J. C. 2003, ApJ, 599, 1082
\bibitem[Majewski et al.(2011)]{MZN11} Majewski, S. R., Zasowski, G., \& Nidever, D. L. 2011, ApJ, 739, 25
\bibitem[Newberg et al.(2002)]{newberg02} Newberg, H. J., Yanny, B., Rockosi, C., et al. 2002, ApJ, 569, 245
\bibitem[Newberg et al.(2007)]{newberg07} Newberg, H. J., Yanny, B., Cole, N., Beers, T. C., Re Fiorentin, P., Schneider, D. P., \& Wilhelm, R. 2007, ApJ, 668, 221
\bibitem[Prior et al.(2009)]{prior09} Prior, S. L., Da Costa, G. S., Keller, S. C., \& Murphy, S. J. 2009, ApJ, 691, 306
\bibitem[Samus et al.(1996)]{Samus96} Samus, N. N., Kravtsov, V. V., Pavlov, M. V., et al. 1996, AstL, 22, 239
\bibitem[Schlegel et al.(1998)]{SFD98} Schlegel, D. J., Finkbeiner, D. P., \& Davis, M. 1998, ApJ, 500, 525
 \bibitem[Vivas et al.(2001)]{vivas01} Vivas, A. K., Zinn, R., Andrews, P., 
 et al. 2001, ApJ, 554, 33
\bibitem[Vivas et al.(2004)]{vivas04} Vivas, A. K., Zinn, R., Abad, C., et al. 2004, AJ, 127, 1158
\bibitem[Vivas \& Zinn(2006)]{vivaszinn06} Vivas, A. K., \& Zinn, R. 2006, AJ, 132, 714
\bibitem[Watkins et al.(2009)]{watkins09} Watkins, L. L., Evans, N. W., Belokurov, V., et al. 2009, MNRAS, 398, 1757
\bibitem[Yanny et al.(2003)]{yanny03} Yanny, B., Newberg, H. J., Grebel, E. K., et al. 2003, ApJ, 588, 824
\bibitem[Zinn et al.(2004)]{zinn04} Zinn, R., Vivas, A. K., Gallart, C., \& Winnick, R. 2004, in ASP Conf.\ Ser., 327, 92
\end{thebibliography}
\end{document}